\documentstyle[11pt,newpasp,twoside,epsf]{article}
\begin{document}
\title{Soft X-ray Spectroscopy of Seyfert~2 Galaxies}
  \author{A. Kinkhabwala, M. Sako, E. Behar, F. Paerels, S.~M. Kahn}
    \affil{Columbia University, 550 W. 120th St., NY, NY 10027}
  \author{A.~C. Brinkman, J.~S. Kaastra, R. van der Meer}
    \affil{SRON, Sorbonnelaan 2, 3548 CA, Utrecht, The Netherlands}
  \author{D.~A. Liedahl}
    \affil{Physics Department, LLNL, P.O. Box 808, L-41, Livermore, CA 94550}

\begin{abstract}
Soft X-ray spectroscopy of Seyfert~2 galaxies provides perhaps the best method 
to probe the possible connection between {\small AGN} activity and star formation.  
Obscuration of powerful radiation from the inferred nucleus allows for 
detailed study of circumnuclear emission regions.  And soft X-ray 
spectroscopy of these regions allows for robust discrimination between
warm gas radiatively-driven by the {\small AGN} and hot collisionally-driven 
gas possibly associated with star formation.  A simple model of a {(bi-)cone} 
of gas photoionized and photoexcited by a nuclear power-law continuum is 
sufficient to explain the soft X-ray spectra of all Seyfert~2 galaxies so 
far observed by the {\it XMM-Newton} and {\it Chandra} satellites.  An upper 
limit of $\sim$10\% to an additional hot, collisionally-driven 
gas contribution to the soft X-ray regime appears to hold for five different 
Seyfert~2 galaxies, placing interesting constraints on circumnuclear star
formation.
\end{abstract}

\section{Introduction}

Emission from warm, recombining gas has been shown to be a key feature of the 
soft X-ray emission from Seyfert~2 galaxies (Sako et al. 2000, 
Ogle et al. 2000, Sambruna et al. 2001).
In Fig.~1, we show the {\it XMM-Newton} Reflection Grating Spectrometer 
({\small RGS}) 
spectrum of the brightest Seyfert~2 galaxy, {\small NGC}~1068 (Kinkhabwala et al. 
2001a).  The presence of multiple radiative recombination continua 
({\small RRC}) 
provide definitive evidence for a dominant recombining gas component with 
temperature $T\simeq3$--5 eV.
Recombination alone, however, is insufficient to explain all of the observed 
X-ray emission from these objects.

\section{Model of Radiatively-Driven Gas Cone}

We propose a model for the X-ray emission from Seyfert~2 galaxies consisting 
of a 
{(bi-)cone} of gas irradiated by a nuclear power-law continuum.  
(The nucleus -- located at the tip of the cone -- is obscured along our 
particular line of sight.)  We specify a single ionic column density, 
$N_{ion}$, and 
gaussian velocity distribution, $\sigma_v$, both {\it along} the cone.  
$\sigma_v$ is 
chosen to be consistent with observations of broadened line absorption in soft 
X-ray observations of Seyfert~1 galaxies, where values from $\sim$200 km/s 
(Kaastra et al. 2000) to $\sim600$ km/s (Sako et al. 2001) have been found.  
We note that $\sigma_v$ may be due to a superposition of multiple velocity 
components.  A broadening of the observed lines due to a separate gaussian 
velocity distribution along our particular line of sight (probably unrelated 
to the velocity distribution along the cone) is also taken.

\begin{figure}
\plotfiddle{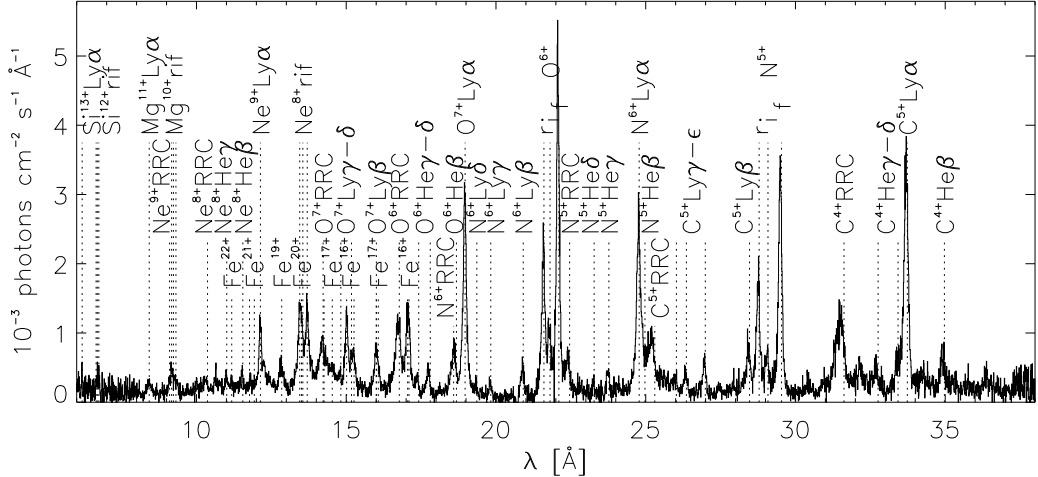}{2.in}{90}{60}{60}{235}{-93}
\caption{{\it XMM-Newton} {\small RGS} spectrum of {\small NGC}~1068.  Note the bright 
{\small RRC},
strong forbidden lines in the helium-like triplets of oxygen and nitrogen,
and relatively-strong higher-order transitions (labelled $\beta$, $\gamma$, 
$\delta$, etc. up to the {\small RRC}).  All of which are unambiguous 
signatures of warm photoionized and photoexcited gas (Sako et al. 2000).}
\end{figure}

\begin{figure}
\plotfiddle{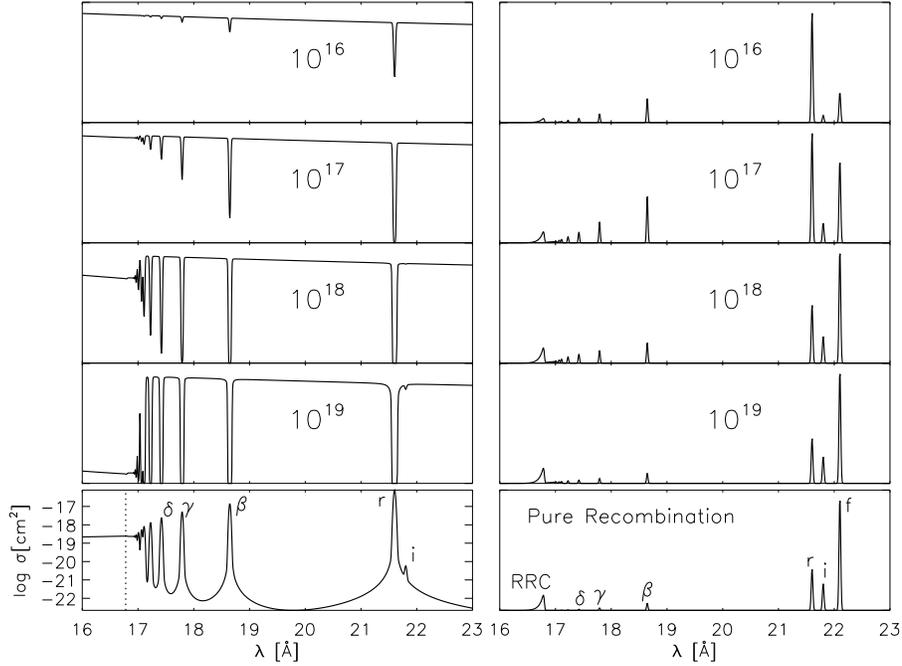}{3in}{90}{50}{50}{200}{-35}
\caption{Effect of varying column density ($N_{ion}=10^{16}$ -- $10^{19}$ 
cm$^{-2}$) along the cone to absorbed (``Seyfert~1'' view on the left) and 
reemitted (``Seyfert~2'' view on the right) spectra for O~{\small VII}.  For 
``Seyfert~2'' view, note the varying relative strength of resonant transitions 
to pure recombination emission (bottom right panel).  ($\sigma_v=200$ km/s 
with linear, but arbitrary vertical scales for flux.)  
Bottom left panel gives cross section for photoexcitation/photoionization with 
separating boundary.}
\end{figure}

\begin{figure}
\plotfiddle{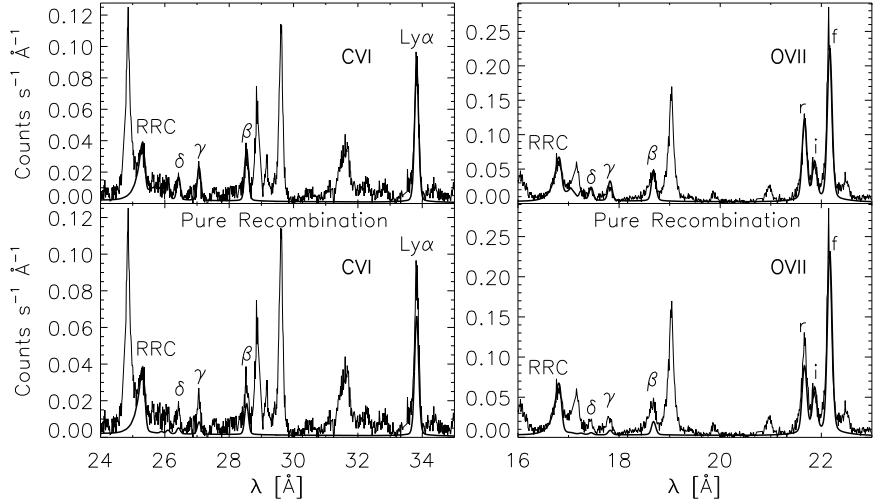}{2.15in}{90}{50}{50}{200}{-90}
\caption{Fits to C~{\small VI} and O~{\small VII} for {\small RGS} spectrum of 
{\small NGC}~1068 ($\sigma_v=200$ 
km/s along the cone and $N_{ion}=10^{18}$ for both ions).  Bottom panels show 
recombination alone, demonstrating presence of significant photoexcitation.  
For C~{\small VI} fit, a factor-of-two reduction of photoionization relative 
to photoexcitation was taken for a better fit (possibly due to 
absorption by N~{\small VII} Ly$\alpha$ shortward of the C~{\small VI} {\small RRC}).}
\end{figure}

The upper right panel in Fig.~2 shows the expected spectrum of the helium-like 
O~{\small VII} ion.  The panels on the left show the ``Seyfert~1'' view down 
the axis of 
the cone in absorption, and the panels on the right show the ``Seyfert~2'' view
in reemission for column densities in O~{\small VII} from $10^{16}$ to 
$10^{19}$ cm$^{-2}$.  All photons absorbed out of the power-law continuum in 
the left panel are reprocessed and reemitted in the right panel.  Further 
details of our model can be found in Kinkhabwala et al. (2001a) and Behar et 
al. (2001).

\begin{figure}
\plotfiddle{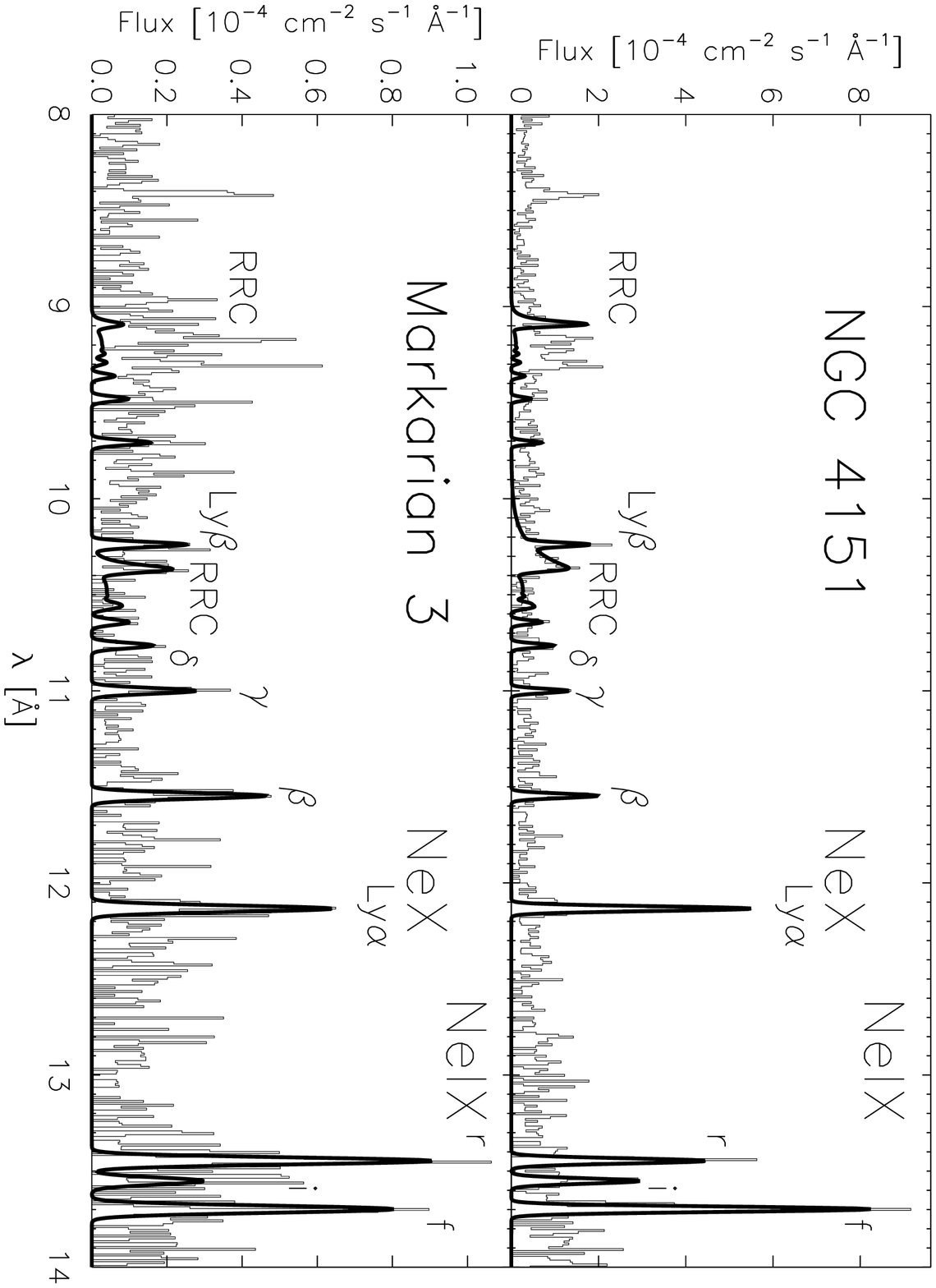}{3in}{90}{50}{50}{200}{-40}
\caption{{\it Chandra} {\small HETG} spectra of {\small NGC}~4151 and 
Markarian~3.  Spectral differences are mostly due to differing Ne~{\small IX} 
and Ne~{\small X} column densities (e. g., Fig. 2) of $2\times10^{18}$ and 
$3\times10^{18}$ cm$^{-2}$ for {\small NGC}~4151 and $5\times10^{17}$ and 
$1\times10^{18}$ cm$^{-2}$ for Markarian~3 (using $\sigma_v=200$ km/s).
}
\end{figure}

\section{Model Fits to Seyfert~2 Galaxy Spectra}

Our model works remarkably well for explaining the bulk (and possibly all) of 
the X-ray emission from {\small NGC}~1068.  Fig.~3 shows our fit to lines
associated with transitions in O~{\small VII} for the {\small RGS} spectrum 
shown in Fig.~1.  Similarly, in 
Fig.~4, we show fits to the previously published {\it Chandra} {\small HETG} 
spectra 
of two other Seyfert~2 galaxies.  We find that the claimed observation of hot 
collisionally-driven gas in {\small NGC}~4151 of Ogle et al. (2000) was 
premature, and we 
verify the earlier conclusions for Markarian~3 of Sako et al. (2000), who 
claimed that photoionization and photoexcitation in addition to a scattered 
power-law continuum were sufficient to explain its observed X-ray spectrum.

We estimate an upper limit of $\sim$10\% of the soft X-ray 
emission may be due to hot, collisionally-driven gas in {\small NGC}~1068.  
Preliminary analysis of 
{\it Chandra} {\small HETG} spectra of four other Seyfert~2 galaxies 
(Markarian~3, {\small NGC}~4151, Circinus, {\small NGC}~4507) suggests that 
their remarkably similar spectra are 
also dominated by reprocessed {\small AGN} emission, with a similar upper limit to a 
hot collisional gas component.
This places interesting limits on the amount of star formation in their
circumnuclear environments (Kinkhabwala et al. 2001a,b).

\end{document}